**A contactless scanning near-field optical dilatometer imaging the thermal expansivity of inhomogeneous 2D materials and thin films at the nanoscale.**

*Victor Wong, Sabastine Ezugwu & Giovanni Fanchini\**

V. Wong, S. Ezugwu & G. Fanchini
Department of Physics & Astronomy, University of Western Ontario, London ON, Canada N6A3K7
E-mail: gfanchin@uwo.ca

G. Fanchini
Department of Chemistry, University of Western Ontario, London ON, Canada N6A5B7

Keywords: Thermal Expansion Coefficient, Scanning-near field optical microscopy (SNOM), Thermoreflecance, Nanogranular materials.

To date, there are very few experimental techniques, if any, that are suitable for the purpose of acquiring, with nanoscale lateral resolution, quantitative maps of the thermal expansivity of 2D materials and thin films, despite huge demand for nanoscale thermal management, for example in designing integrated circuitry for power electronics. Besides, contactless analytical tools for determining the thermal expansion coefficient (TEC) are highly desirable, because probes in contact with the sample significantly perturb any thermal measurements. Here, we introduce $\omega$-$2\omega$ near-field thermoreflectance imaging, as an all-optical and contactless approach to map the TEC at the nanoscale with precision. Testing of our technique is performed on nanogranular films of gold and multilayer graphene (ML-G) platelets. Our method demonstrates that the TEC of Au is higher at the metal-insulator interface, with an average of $(17.12\pm2.30)\times10^{-6}$ K$^{-1}$ in agreement with macroscopic techniques. For ML-G, the average TEC was $(-5.77\pm3.79)\times10^{-6}$ K$^{-1}$ and is assigned to in-plane vibrational bending modes. A vibrational-thermal transition from graphene to graphite is observed, where the TEC becomes positive as the ML thickness increases. Our nanoscale method demonstrates results in excellent agreement with its macroscopic counterparts, as well as superior capabilities to probe 2D materials and interfaces.

## 1. Introduction

With critical advances in integrated power electronics, [1] there has been a tremendous demand for nanoscale thermal management in the pursuit of miniaturization and integration of submicrometric components while preserving high power density for applications in communication, aircraft engines and renewable and sustainable energy storage.[2-4] The need for heat dissipation is essential for many of such applications[5] as it assists in preserving the efficiency and lifespan of the circuitry components. Heat spreaders and heat sinks are typically integrated for the purpose of discharging local thermal buildup from long device operation time.[6] With tremendous demand for miniaturization, understanding of heat dissipation in low-dimensional systems and thin film structures beyond the macroscopic level is rapidly becoming essential towards the development of next generation information technology. However, a major cause of powered electronics failure is due to thermal stress induced by a difference in thermal expansion coefficients of different materials in contact upon long durations of heat exposure.[7] Despite the demand, there are few experimental techniques suitable for the purpose of providing thermal expansion information beyond the macroscopic level.

Thermal expansion of inhomogeneous thin films is customarily investigated macroscopically with dilatometers, [8-10] thus only measuring their average expansivity from bulk materials with known geometry through an external heat source.[8] Microscopic and nanoscopic thermal expansion information is generally not provided by macroscopic dilatometry techniques. In principle, thermal expansivity information with higher resolution can be obtained by measuring the changes in the crystallographic lattice sites with x-ray spectroscopy where the measured spectra are accompanied with a thermodynamical interpretation involving anharmonic effects to determine the thermal expansion.[11-14] These methods are also macroscopic in nature, however, and they require crystalline materials.

To date, guesses on the local thermal properties of surfaces and near-surface features at the mesoscopic level chiefly rely on scanning thermal microscopy (SThM)[15-19] a contact-mode scanning probe microscopy. In SThM, heat is generated in the sample due to an external AC electrical source, while detection is measured through periodical oscillations of the probe cantilever to determine the thermal expansion coefficient of the material. However, heat detection in conventional SThM requires the cantilever to be in contact with the sample surface, which acts as a huge thermal sink that is also irreproducible due to unpredictable sample-probe distance. Furthermore, the cantilever also expands due to heat, thus further complicating the genuine identification of thermal expansion from the sample.[20-22] Alternative methods of nanoscale thermal detection have been proposed,[23-26] and they involve the combination of conventional SThM with other techniques or uses synchrotron-based methods assisted with high resolution transmission electron microscopy.[27] Such methods were also occasionally proposed for acquiring thermal expansion information with nanoscale resolution.[28,29] However, the configurations of the proposed systems are highly specialized and require a combination of multiple techniques, some of which are available only at large-scale synchrotron facilities.

In this paper, we present and utilize for the first time contactless near-field thermo-reflectance imaging, a novel tool for precisely mapping the thermal expansion coefficient at the nanoscale. Near-field thermoreflectance[30,31] is a class of pump-probe techniques based on super-resolution, aperture-type, scanning near-field optical microscopes (SNOMs) –optical instruments capable of lateral resolution beyond the diffraction limit through the use of evanescent light waves.[32,33] Aperture-type SNOMs make use of atomic-force cantilevers in which nano-sized apertures are drilled via focussed ion-beams and, therefore, the acquisition of an atomic force micrograph (AFM) of the sample surface is possible in conjunction with nano-optical imaging.[34-38] In pump-probe near-field thermoreflectance experiments, samples are periodically heated by a "pump" laser beam from an inverted optical microscope, while periodic changes in the sample's local reflectivity caused by the resulting heat waves are measured, in phase and amplitude, through an upright aperture-type SNOM operating in reflection mode from a low intensity laser ("probe") of wavelength different from the "pump". Near-field thermoreflectance tools can be operated in non-contact mode,[30,31] thus making them contactless and offering significant advantages, in terms of little invasiveness, over methods requiring electrical currents to heat the sample, or nanosized thermocouples to measure their thermal properties, or both.

The use of scanning near-field thermoreflectance methods to map the thermal conductivity and heat capacity of specific samples, by locking the probe beam detection at the same frequency $\omega$ as the pump beam in an $\omega$-$\omega$ pump-probe technique, is well assessed.[30,31] Here, we demonstrate that one-of-a-kind $\omega$-$2\omega$ near-field thermoreflectance experiments, in which the probe is locked-in for detection at double frequency ($2\omega$) as the pump, are spectacularly sensitive to the local expansivity of surfaces, and can thus be successfully used to satisfy the longstanding quest for a contactless tool capable of imaging the thermal expansion coefficient (TEC) of composite materials at the nanoscale. This novel technique, which can be termed scanning near-field optical dilatometry (SNOD) is here demonstrated to quantitatively estimate and successfully image the nanoscale thermal expansion coefficient of samples with both positive and negative thermal expansivity. SNOD is anticipated to be uniquely suited for thermal expansion coefficient imaging in a variety of thin films and devices, thus solving longstanding fundamental and technological problems in nanoscale thermal management.

## 2. Results and Discussion

### 2.1. Operation Principles of SNOD

Panels a and b in **Figure 1** elucidate the principle of operation of thermoreflectance experiments and the differences between $\omega$-$\omega$ and $\omega$-$2\omega$ operation modes, where this second mode of operation is specifically pioneered in this work. Samples include a transparent substrate with an optically absorbing and heterogeneous thin film on the top. As these thin films are heterogeneous and of different material properties, their TEC is expected to vary from point to point (x,y) of the probed surface. Due to the high transparency of the sample, the pump beam ($\lambda_0$=405 nm, purple, in figure 1a-b) pulsed at a frequency $\omega$ through an optical chopper and periodically heats the film, from which heat dissipates both along the surface or through the substrate and, to a limited extent, air at the top. Air experiences a stronger change in density than the solid sample, which results in oscillations of the air refractive index and surface reflectivity [$\Delta\rho_\omega(x,y)$]. Such oscillations are probed by an upright SNOM in reflection mode ($\lambda$=532 nm in figure 1a-b). This allows to locally probe the oscillations in film's surface temperature [$\Delta T_{f,\omega}(x,y)$] which are linear in the reflectivity:

$$\Delta T_{f,\omega}(x,y) = \Delta\rho_\omega(x,y) / h, \qquad (1)$$

where $h$ is a constant that only depends on $\lambda$, on the material, tip aperture radius and length, as well as the focal distance from the detector.[30]

The interpretation of ω-ω measurements summarized by eq 1 assumes that the near-field thermo-reflectance entirely originates from the sample surface, which on its turn assumes the evanescent field to be spatially localized at lengths much shorter than the spatial fluctuations of the sample surface along the z-axis. [30] This is normally not the case, as the near-field decays over lengths of the order as the tip radius *a* (i.e. a few tens nm) and these fluctuations can be either time-independent [i.e., from position dependent $z = z(x,y)$] or time-dependent, from $z = z(t)$ at a given scanned point (x,y). Performing SNOM measurements in AFM non-contact mode eliminates the contribution from time-independent fluctuations, but not from time-dependent ones, as the tip resonant frequency is normally much larger than ω. As seen in figure 1b, the thermal expansivity [$\varepsilon(x,y)$] of the sample surface under sinusoidal heating induces time-dependent oscillations of the z-axis position of the sample surface at the same frequency:

$$z(x,y,t) = z_0 + \varepsilon(x,y)\, cos(\omega t). \qquad (2)$$

Because the reflected evanescent field is not point-like, but can be approximated as a gaussian profile peaked at a constant distance $z_0$ from the sample, and of full width at half maximum of about the tip aperture radius *a*, the near-field reflectivity from temperature-induced, time-dependent fluctuations of the underlying surface is given by:

$$\Delta\rho = \Delta\rho_\omega(x,y)\, exp\{-[z(x,y,t)-z_0]^2/a^2\} \approx \Delta\rho_\omega(x,y)\{1-\varepsilon(x,y)^2[1+cos(2\omega t)]/2a^2\}, \qquad (3)$$

where a Taylor expansion of the gaussian can be used to work out the right-hand term of eq 3 in conjunction with eq 2 and the half-angle formula $2cos^2(\omega t) = 1 + cos(2\omega t)$. Thus, the amplitude of the time-dependent SNOM signal at double-frequency (2ω) is given by:

$$\Delta\rho_{2\omega}(x,y) = \Delta\rho_\omega(x,y)\, \varepsilon(x,y)^2 / 2a^2. \qquad (4)$$

$\varepsilon(x,y)$ can be derived from $\Delta\rho_{2\omega}$ by means of eq 4, if $\Delta\rho_\omega(x,y)$ is known from an independent ω-ω experiment, and the lateral resolution of this method will be ~ $a$. The z-axis resolution will be more accurate, however, as the use of a lock-in method enables the detection of quantities $\varepsilon(x,y) / a \sim 10^{-2}$ and, therefore $\varepsilon(x,y) \approx 0.5$ nm with $a \approx 50$ nm aperture. As the two experiments ($\omega$-$\omega$ and $\omega$-$2\omega$) can be carried out simultaneously from the same instrumentation, this finding opens a robust pathway towards contactless near-field based scanning optical nanodilatometry.

An immediate way to picture the underlying principle of contactless SNOD through $\omega$-$2\omega$ experiments is depicted in figure 1b. Consider the sample surface to be, on average, at temperature and height $T_{Mid}$ and $z_{Mid}$, respectively. At a time $t_{Hi} = \pi/2\omega$ during a pump-beam cycle, the sample is heated up to $T_{Hi}=T_{Mid}+\Delta T$, which expands it locally, thus placing its surface at $z_{Hi} = z_{Mid}-\varepsilon(x,y)/2$ below the tip. Cooling the sample at a time $t_{Lo} = 3\pi/2\omega$ during the same cycle, will bring the temperature down to $T_{Lo}=T_{Mid}-\Delta T$, thus locally placing its surface at a larger distance, $z_{Hi} = z_{Mid}+\varepsilon(x,y)/2$, from the tip. Hence, the temperature and position of the sample surface relative to the contactless tip oscillate at $\omega$. However, the sample surface sits away from the gaussian peak of the reflected evanescent field at both $t_{Hi}$ and $t_{Lo}$, where the sample reflectivity reaches equal minima, $\rho_{Lo} = \rho_{Lo} = \Delta\rho_\omega-\Delta\rho_{2\omega} = (1-\varepsilon^2/2a^2)\Delta\rho_\omega$ because of the symmetry of the gaussian-shaped profile of the reflected evanescent field. Because the reflectivity reaches two minima, at $\pi/2\omega$ and $3\pi/2\omega$, for each pump-beam period, then the reflectivity signal from thermal expansivity oscillations occurs at double frequency as the excitation. In addition, the amplitude of the reflected evanescent field will reach maxima at $t_{Mid} = 0$, $\pi/\omega$, and $2\pi/\omega$, also twice as frequently as the pump beam, thus confirming the $2\omega$ oscillations of the signal associated to the sample's thermal expansion.

After measuring $\varepsilon(x_0,y_0)$ via eq 4, the next critical step is represented by properly taking into account the contributions from the substrate on the determination of the local TEC [$\alpha_f(x_0,y_0)$] of the film, where the former can be substantial (up to 50%) even for thermally insulating substrates like glass. The two quantities are linked by the relationship: [8-10]

$$\varepsilon(x_0,y_0) = \alpha_f(x_0,y_0)\, d(x_0,y_0)\, \Delta T_f(x_0,y_0) + \alpha_s \int dz_0\, \Delta T_s(x_0,y_0,z_0), \qquad (5)$$

where $\Delta T_f(x_0,y_0)$ can be obtained through $\omega$-$\omega$ thermoreflectance measurements via eq (1), and $d(x_0,y_0)$ is the local thin-film thickness. Being scanning-probe techniques, SNOM and SNOD are uniquely suited to determine $d(x_0,y_0)$ through a simultaneous AFM scan of the sample. Also, in eq 5, $\alpha_s$ is the TEC of the substrate, which, for most of the cases of practical interest, is known and expected to be homogeneous. More complicate is the determination of $\Delta T_s(x_0,y_0,z_0)$, the change in temperature at a point of the substrate at a level $z_0$ below the measured point of the surface. Multiple neighboring substrate points (x,y) may affect the thermal expansion around $(x_0,y_0)$. In an ideal case where the substrate is perfectly insulating, and the film is enough conducting for heat to be overwhelmingly dissipated in-plane along the surface, the integral from the second addend of eq 5 yields a negligible value, but this is normally not the case. [30]

To facilitate the determination of eq 5, $\Delta T_s(x_0,y_0,z_0)$ will be divided in two components, as depicted in figure 1c: i) an (x-y)-independent component [$T_{s,av}(z_0)$] originating from the film's temperature oscillation $T_{f,av}$ averaged over the entire scanned region (panel c, top), and ii) the local deviation from the average [$\delta T_s(x_0,y_0,z_0)$] which originates from the fluctuations of the measured film's temperature $\delta T_f(x_0 y_0)$ (panel c, bottom). Due to the linearity of the equation of heat, this can be solved independently for cases i) and ii), by using $T_{s,av}(0)=T_{f,av}$ and $\delta T_s(x_0,y_0,0)=\delta T_f(x_0 y_0)$, respectively, as boundary conditions, and the solutions can be superimposed as

$$\Delta T_s(x_0,y_0,z_0) = T_{s,av}(z_0) + \delta T_s(x_0,y_0,z_0), \tag{6}$$

where

$$\delta T_s(x_0,y_0,z_0) = \iint_A dx\, dy\, \delta\Theta_s(x-x_0,y-y_0,z_0), \tag{7}$$

and A is the considered cut-off area of the sample. $\Delta T_s(x_0,y_0,z_0)$ obtained from eq 6 can then be integrated over $z_0$ in eq 5 to yield $\alpha_f(x_0,y_0)$. Specifically, component i) offers a substrate temperature uniformly decaying over z irrespective of (x,y) (figure 1c, top), while component ii) results in a local temperature oscillation of the form $\delta\Theta_s(x-x_0,y-y_0,z_0)$, which is the product of a spherical harmonics that only depends on $r = [(x-x_0)^2 + (y-y_0)^2 + z_0^2]^{1/2}$ and an azimuthal component only depending on $\varphi = atan\{z_0/[(x-x_0)^2+(y-y_0)^2]^{1/2}\}$ (figure 1c, bottom, see also Supporting Information). Under the additional assumption that multiple $\delta\Theta_s(x-x_0,y-y_0,z_0)$ from neighbouring substrate points have random signs, and tend to cancel out due to the randomness of thermal fluctuations in disordered media, the integration in eq 5 is, at the present stage, carried out by neglecting the contributions from any component at $x \neq x_0$ and $y \neq y_0$ (i.e. with a cut-off area A corresponding to 1 pixel, so $\delta T_s(x_0,y_0,z_0) \approx \delta\Theta_s(x_0,y_0,z_0)$). Nonetheless, it is anticipated that this simplified model of the substrate's contributions to the thermal expansion is sufficient to describe optically transparent substrates at low enough thermal conductivity.

In the following section, we explore the use of SNOD for investigating specific systems of practical interest, with both positive and negative TEC. The objective is to test our method with special attention to nanogranular and sparse films (where macroscopic dilatometry techniques may be inadequate) as well as highly thermally conducting systems such as gold and graphene, where the use of SThM is problematic due to strong heat transfer to the probe.[20-22]

## 2.2. Demonstration of SNOD imaging

*2.2.1. Nanogranular and continuous thin films of gold on glass*

Nanogranular but continuous thin films of gold on glass are ideal testing samples for SNOD as their data interpretation is simplified by the fact of heat being predominantly dissipated along the Au thin-film surface, which is very thermally conducting, while the relative uniformity of the thin film minimizes $\delta T_f(x_0,y_0)$ from figure 1c. For this system, the substrate's contribution to the sample's expansivity from eq 5 is overwhelmingly represented by $T_{s,av}(z_0)$ in eq 6, which is independent of $(x,y)$ and, consequently $(x_0,y_0)$.

**Figure 2** shows the SNOD images and the corresponding calculated thermal expansion maps from an 80-nm thick Au film, where the sample granularity is noticeable from the AFM micrograph in panel a, with SNOM measurements in transmission and reflection mode in panels b-c, respectively. The corresponding $\omega$-$\omega$ thermoreflectance map is shown in panel d, with their $\omega$-$2\omega$ thermal expansion counterparts presented in panels e-h. The TEC is an intensive property that, in the bulk of a solid, will only depend on the crystal structure; even though it can assume different values at surfaces and interfaces,[39] it cannot depend on measuring parameters, such as $\omega$. To validate this, thermal expansion maps are measured for $\omega$ = 45, 60, 80 and 100 Hz in the same region as the AFM micrograph in panel a. We can thus observe that the thermal expansion maps are independent of $\omega$, which is an indication of their genuineness. However, it can be observed that the TEC is highest in the proximity of Au-glass interfaces (panel i) and relatively constant elsewhere. From panels e-h, we estimate the average volumetric TEC of Au to be (51.35 ± 6.89) x $10^{-6}$ K$^{-1}$ with an average linear TEC of (17.12 ± 2.30) x $10^{-6}$ K$^{-1}$, and uncertainties determined by the standard deviation from all the measured ω. Comparison of these values with the TEC for Au recorded by macroscopic dilatometry techniques from the literature[40-54] is reported in **Table I**. In addition, the TEC of the same nanogranular Au thin film probed by SNOD was also measured macroscopically using an optical dilatometer (see Supporting Information) and comparable TEC estimates were obtained. We can thus infer that SNOD provides quantitatively correct measurements of TEC of Au thin films, with additional information at the grain boundaries and interfaces with the glass substrate.

*2.2.2. Sparse multilayer graphene (ML-G) platelets on glass*

To further substantiate the capabilities of SNOD, we have performed the same measurements on a system of complementary characteristics, which is formed by sparse multi-layer graphene (ML-G) platelets deposited on optically transparent and thermally insulating glass,[55] as summarized in **Figure 3.** As far as this system is concerned, the substrate's contribution to the sample's expansivity from eq 5 is overwhelmingly represented by $\delta T_s(x-x_0,y-y_0,z_0)$ in eq 6, which, at a certain subsurface level $z_0$, may depend on a relatively large number of substrate points (x,y) around ($x_0,y_0$). However, because of the sparseness of the ML-G platelets, heat is here overwhelmingly dissipated through the substrate, orthogonally from the surface, and therefore a small cut-off area A in eq 7 appears to be adequate.

From the AFM micrograph in figure 3a, it can be observed that the ML-G platelets are dispersed over the substrate surface with little overlap, and some folding, which results in a broad range of number of stacking layers. The SNOM transmission and reflection maps over the same region are presented in Figure 3b and c respectively, while panel d shows the temperature profile of the sample. A local optical absorption profile can be determined using the near-field transmission and reflectance signals to determine the local temperature profile and be used to calculate the local thermal expansion maps from $\alpha_f(x_0,y_0)$ in eq 5. TEC maps of ML-G platelets are presented in figure 3e-i for frequencies $\omega$ = 100, 200, 400 and 700 Hz, respectively. Because the TEC is a material specific-parameter, the results in figure 3e-i should not depend on $\omega$ and any differences are the result of measuring uncertainties. In fact, the differences between the TEC determined from such independent measurements are marginal, and consistent in magnitude with normal fluctuations during sequential AFM and SNOM scans.

From the summary in figure 3h, which correlates average TEC results from panels e-i and the ML-G thickness from AFM (panel a), we can observe that the thinner ML-G the more negative the TEC. A transition point can be observed at approximately 175 layers. For platelets thicker than 175 layers, the TEC transitions to positive, as it happens in bulk graphite.[57] Furthermore, the linear TEC above 175 layers undergo a plateau at $(33.75 \pm 3.24) \times 10^{-6}$ K$^{-1}$ which is consistent with the out-of-plane thermal expansion of graphite along the crystallographic c-axis,[56-59] which corroborates the validity of SNOD as an accurate and quantitative method for thermal expansivity measurements. Conversely, a value of $(-5.77 \pm 3.79) \times 10^{-6}$ K$^{-1}$ is observed via SNOD for the thinnest ML-G platelets, below about 90 layers, which can be assigned to the in-plane bending modes of graphene,[56,57] and is consistent with previous macroscopic measurements,[57-70] as shown in **Table II**. A significant advantage of SNOD over all of such techniques rests in that it provides site-specific thermal expansion information while other probing methods only yield an ensemble average of the thermally induced expansion and provides minimal information on the distribution of the thermal expansion throughout the whole sample. In addition, through the systematic use of newly invented SNOD technique, we have been able to prove that the transition between graphene and graphite occurs, from a thermal expansion point of view, at about 175 layers and, therefore, at a higher thickness than the transition of electronic properties between few-layer graphene and graphite, which typically occurs at about 10 layers or below.[71]

## 3. Conclusions

In this article, we have described the invention and construction of the first contactless nano-dilatometry apparatus capable of imaging the TEC at the nanoscale in nanostructured thin-film materials, 2D materials and relating electronic nanodevices. Our apparatus, a scanning near-field dilatometer, is all-optical and, therefore, does not require any external heater or electrical contact to vary the sample's temperature, nor it requires any physical thermometer or nano-thermocouple to probe it. Our method is a $\omega$-$2\omega$ pump-probe method, where the probe is an aperture-type reflection-mode SNOM, which makes it uniquely suited for probing optically absorbing and thermally conducting layers on optically transparent and relatively thermally insulating substrates, albeit the role of the substrate in determining the measured thermal expansion is considered, through a rigorous model based on Fourier's heat equation.

Using SNOD, we also have found that the TEC of nanogranular Au films is shown to be higher at the glass/Au interface. Such a finding would have been impossible with any other techniques. It would have been impossible with macroscopic contactless techniques because of poor resolution, which would be insufficient to resolve them by several orders of magnitude. To the best of our knowledge, the same effect has never been observed before using SThM, arguably because the strong interaction of supposed SThM "nano-thermocouples" (that are, in fact, rather voluminous) may produce a too large lateral heat spreads for thermal expansion gradients to be observed over few microns.

As far as 2D materials are concerned, SNOD has led us to find for the first time to observe that, from a thermomechanical point, of view, the graphene-graphite transition occurs at ~175 layers. Such a finding has so far been elusive with other techniques because of the difficulty to reproducibly heat highly thermally conducting ML-G flakes on a thermally insulating samples without a contactless nanoscale dilatometer. The discovery that such a transition occurs over a hundred layers is particularly remarkable because, as far as its electronic structure is concerned, few-layer graphene is known to transition to graphite at about 10 layers or less,[71] Therefore, the discovery that, from a thermomechanical standpoint, "graphene is thicker than we thought" may have critical implications for which include, for example, biodevices and bioimplants, in addition to nano-electronic devices and heat spreaders. Collectively, our work heralds the significance of the development of a near-field based super-resolution optical technique for nanodilatometry imaging applications. Owning to the unique properties of evanescent waves, we can probe thermal events beyond the limits of diffraction which will be invaluable for studying a large variety of 2D materials and nanostructured thin-film systems in which thermomechanics is critical for their applications.

**Materials and Methods**

*SNOM Apparatus and Transmission/Reflection SNOM scans.* Experiments described in this study were recorded on a WiTec Alpha 300S aperture type SNOM/AFM consisting in an AFM oscillator that can equipped with hollow cantilevers (SNOM-NC, NT-MDT) with 20x20x13 µm LxWxH pyramidal tips, at the end of which a subwavelength aperture (typical radius: 60 nm) was obtained by focussed ion beam (FIB) milling. The AFM oscillator is situated at the focal plane of an upright laser confocal optical microscope, from which different lasers can be focused to produce an evanescent field next to the tip aperture. The WiTec Alpha 300S can acquire both transmission-reflection-mode SNOM images while simultaneously measuring the surface morphology via AFM (panels a in figs. 2 and 3). Transmission-mode SNOM images (panel b in figs. 2 and 3) were obtained by illuminating the sample with light focused through the SNOM cantilever with a 405 nm diode laser (500 mW, Apinex) that is passed through a single-mode optical fiber (OZ Optics, NA=0.12). Evanescent waves generated at the apex of the cantilever is used to probe the sample where scattered photons that emerge from the substrate are collected with an inverted optical microscope from below the substrate, and fed into a photomultiplier tube (U64000 Hamamatsu). Reflection-mode SNOM images (panel c in figs. 2 and 3) were obtained within the same apparatus as the transmission-mode scans, but with scattered photons collected at a grazing angle in the near-field by a subminiature accessory (SMA, WiTec) and, from there, fed into the photomultiplier tube. It was essential to be able to acquire both transmission- and reflection-mode SNOM images as they provide information on the local sample absorbance and, from there, determine the sample heat profile in conjunction with knowledge of the pump-beam illumination.

*Thermoreflectance ($\omega$-$\omega$) and Nano-dilatometry ($\omega$-$2\omega$, SNOD) scans*. $\omega$-$\omega$ and $\omega$-$2\omega$ thermo-images (corresponding, respectively, to panel d and panels e-h in figs. 2-3) are recorded using the above mentioned WiTec instrument in the configuration previously reported in **Figure 1**, with the cantilever in noncontact (liftoff) mode. In this configuration, the sample is illuminated ("pumped") with a 405-nm laser beam (500 mW, Apinex) from below the sample through the inverted optical microscope previously used to collect the transmission-mode SNOM signal (see section above). This 405 nm laser is modulated with a mechanical chopper (SciTec Instruments, 300 CD/HRG) and the chopper angular velocity, ω, is varied to record different scans (as in panels e-h of figs. 2-3). During the pulsed illumination and heating of the sample from this 405-nm pump beam, the sample surface is scanned ("probed") in the near field with the upright SNOM microscope. These SNOM measurements are carried out by focussing a 532-nm laser (50 mW Excelsior, Spectra Physics) at the FIB-milled aperture of the scanning SNOM tip (SNOM-NC, NT-MDT, see above) through the upright confocal optical microscope. It is worthwhile noting that the power at the sample of this 532-nm probe beam is significantly lower than the 405-nm pump's. The scattered light off the surface of the sample is collected at a grazing angle by the WiTec SMA reflection mode accessory. A 405 nm notch filter (Thorlabs Inc., NF405-13) and 530 nm long pass filter (Melles-Griot) are positioned in series with the reflection mode SMA to ensure the detected optical signal at the photomultiplier purely originates from the 532-nm probe-beam light, not from the 405-nm pump. The signal from at each point on the sample is obtained by connecting the photomultiplier tube with a dual phase lock-in amplifier (830, Stanford Research Systems) with the reference signal originating from the mechanical chopper, either at $\omega$[30] (panels d of figs. 2-3) or $2\omega$ (panels e-h of figs. 2-3) frequency. Raw image data were saved in matrix format and processed in *ad hoc* Matlab$^{TM}$ routines implementing the model presented by eqs. 1 and 4-7 to obtain the TEC images.

*Testing samples–Nanogranular and continuous thin film of gold on glass.* Nanogranular gold thin films were selected due to the chemically inertness of Au as a noble metal. Au thin films were thermally evaporated onto 1x1-in glass substrates (Corning 7101). Glass substrates were pre-cleaned in a bath sonicator with multiple 15' baths (soap water, deionized water, reagent-grade acetone, and methanol). The substrates were then blown dry with medical-grade nitrogen. The cleaned substrates were loaded into a custom-built thermal evaporator,[39] and growth was monitored through a Sycom STM-2 thickness monitor controlling an Inficon EasyRate water-cooled sensor equipped with a Kurt-Lesker 5-MHz quartz oscillator. A growth rate of 2 nm/min up to 80 nm thickness was used. 80 nm is expected to be the optimal trade-off for SNOD imaging between film thinness (required for homogeneous optical absorption and heat generation along the z-axis) and film thickness (required for substantial heat dissipation in x-y direction to use these as test samples). The quality of the film accordingly to specifications was tested by AFM prior to SNOD measurements.

*Testing samples–Sparse ML-G platelets on glass.* ML-G nanoplatelets on glass substrates (Corning 7101) were prepared accordingly to the method by Sharifi *et al.*[55] Briefly ML-G was obtained from surfactant-assisted exfoliation of nanocrystalline graphite powder (average flake dimension: $d_{av} \approx 700$ nm) exfoliation using RNA VI from *torula utilis* (Sigma-Aldrich) as a surfactant, forming a stable ML-G suspension in water. These suspensions were filtrated on nitrocellulose membranes (EMD, Millipore) which were transferred on the requisite glass substrate and subsequently etched in multiple acetone and methanol baths, leaving behind an ensemble of surfactant-free nanoplatelets on their substrate, which were thoroughly washed in subsequent acetone, methanol and water baths to eliminate any traces of surfactant and filtration membrane, as ascertained by AFM,[55] prior to being submitted for SNOD measurements.

**Acknowledgements**

GF acknowledges a Canada Research Chair in Carbon-based nanomaterials and Nano-optoelectronics and funding from the Natural Sciences and Engineering Research Council (NSERC) of Canada under the Discovery Grant program (506356-2020) and the Canada Foundation for Innovation (LOF-212442). VW acknowledges an NSERC PGS Doctoral Scholarship.

**Figures and Tables**

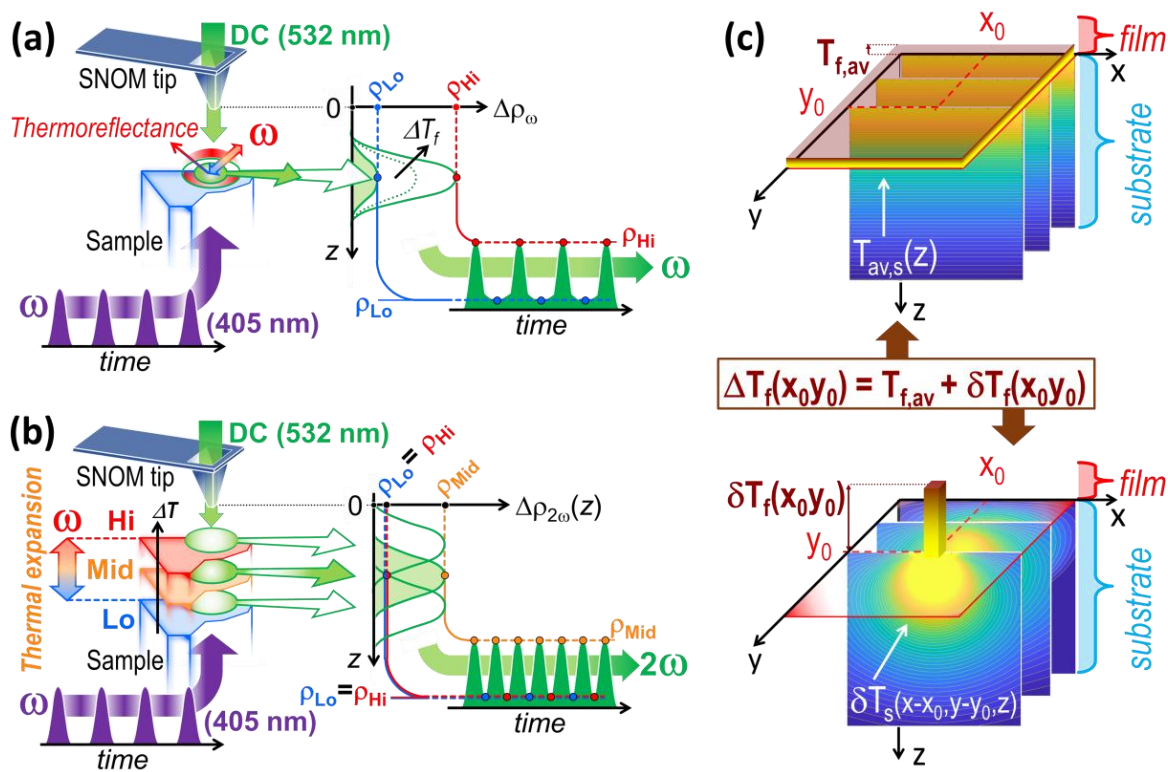

**Figure 1**. a) Schematic of ω-ω pump-probe scanning near-field thermoreflectance imaging experiments for contactless nano-optical probing of temperature oscillations ($\Delta T_f$) at the sample surface via local changes of the air's refractive index; [30] b) ω-2ω pump-probe experiments for contactless nano-optical probing the local thermal expansion of inhomogeneous thin films on transparent substrates (sample). Both experiments can be implemented in a commercial apparatus mounting aperture-type SNOM tips in a scanning probe microscope (AFM) with upright-confocal (532 nm) and inverted (405 nm) optical microscopes for sample illumination, and a submillimeter accessory (SMA) for 532-nm evanescent wave detection in reflection mode. With thermal expansion, the gaussian evanescent wave is off-peak at both maxima (Hi) and minima (Lo) of the sample temperature, so thermal-expansion related oscillations occur at angular frequency 2ω; and c) Model for determining the substrate temperature used in eq 5 from the linear superposition of average ($T_{f,av}$, top) and local ($\delta T_f$, bottom) film temperature oscillations of $\Delta T_f$ as detailed in the Supporting Information.

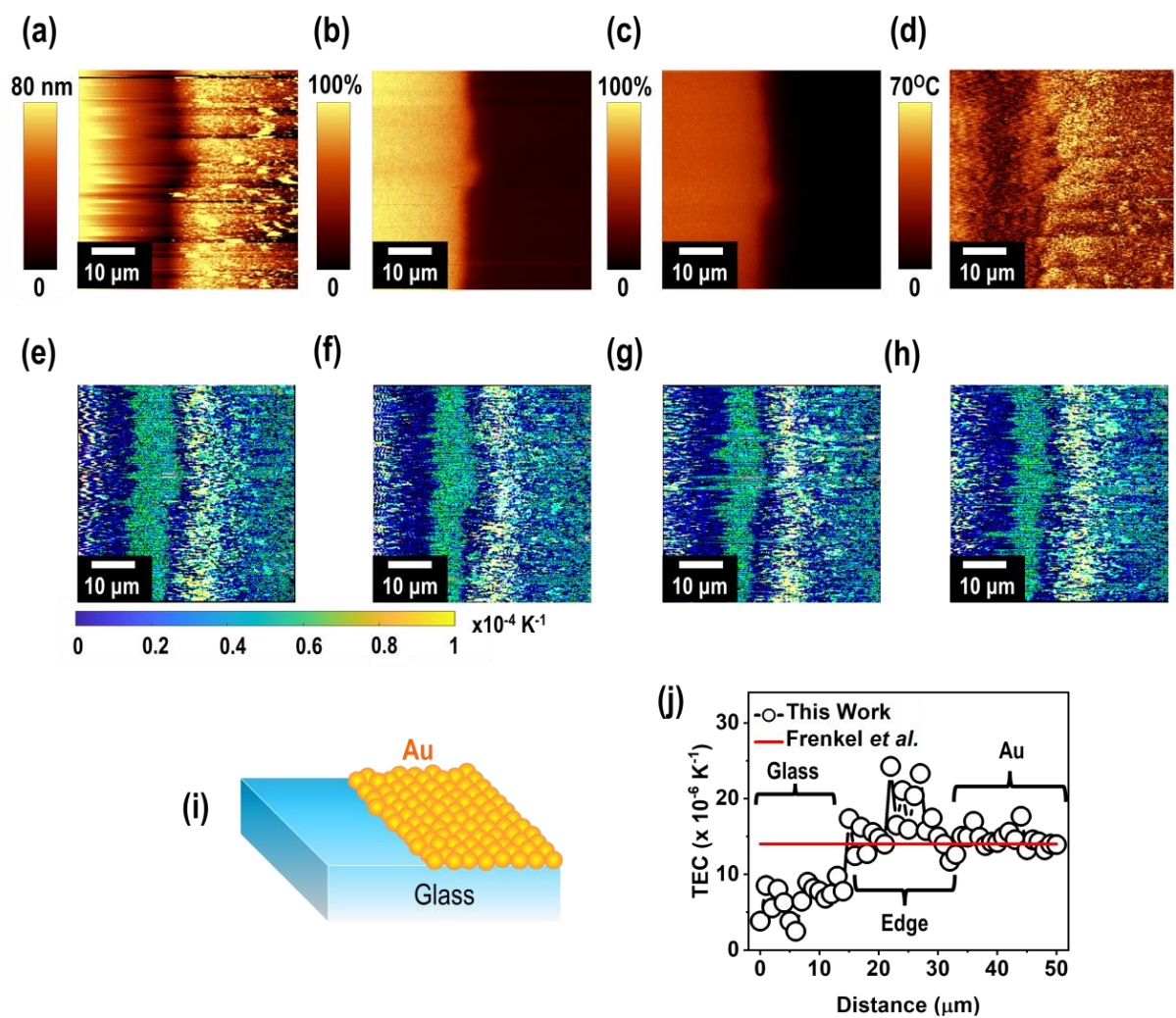

**Figure 2:** a) AFM micrograph of a nanogranular and continuous gold thin film on glass at the metal-glass interface, along with b) transmission-mode and c) reflection-mode SNOM images as well as d) Nanoscale temperature oscillation profile [$\Delta T_f(x,y)$) measured by ω-ω pump-probe near-field scanning thermoreflectance imaging.[30] e) Thermal expansion maps at 45 Hz; f) 60 Hz; g) 80 Hz; and h) 100 Hz obtained from ω-2ω pump-probe measurements. i) Schematic representation of the nanogranular gold thin film, for which j) the linear thermal expansion coefficients (TEC) is observed to be highest at the thin-film edge, corresponding to the metal-

insulator interface, while it is relatively constant and in good agreement with the macroscopic measurements by Frenkel et al[55] at other areas of the nanogranular gold thin film.

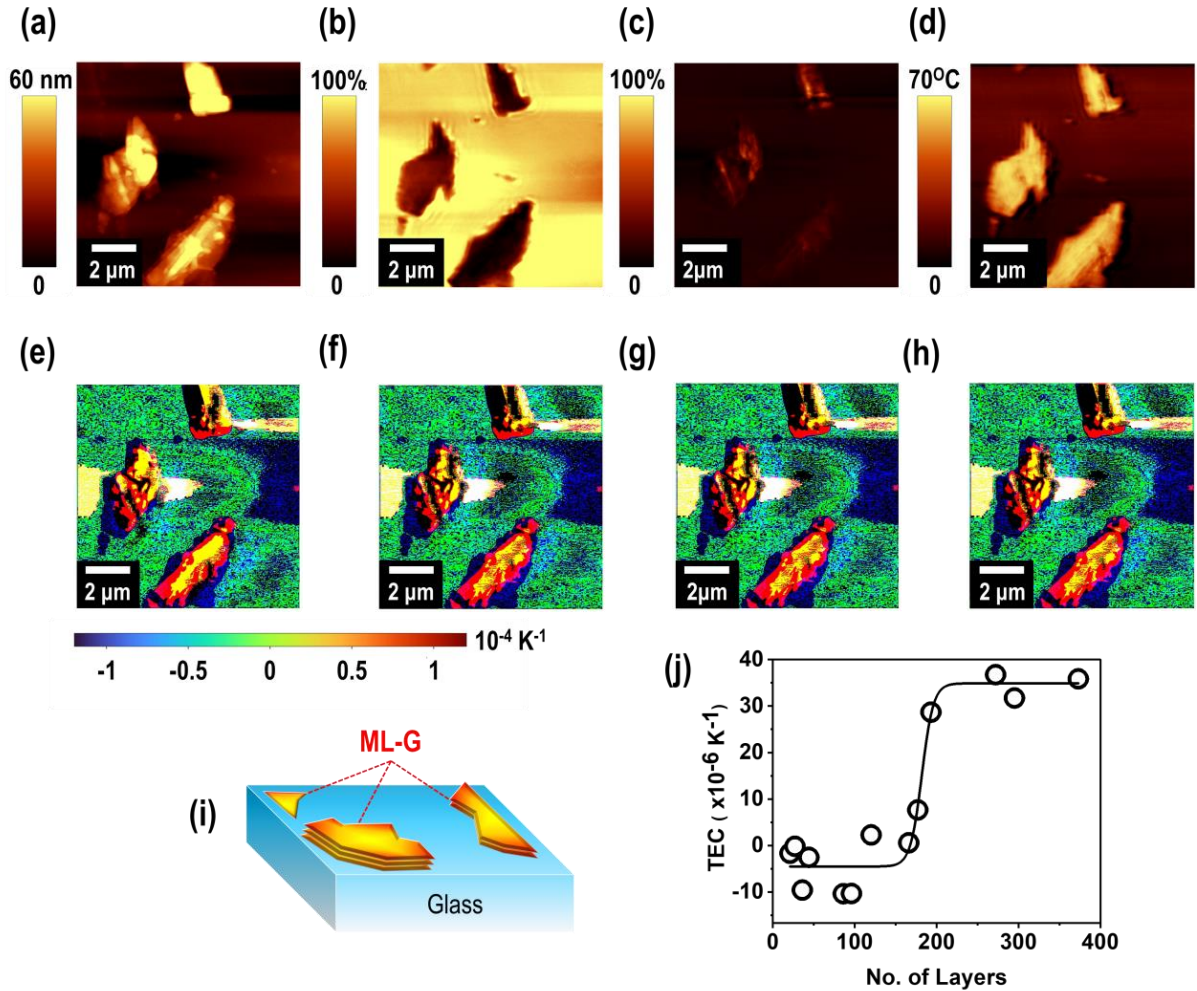

**Figure 3:** a) AFM micrograph of sparse ML-G platelets on glass at the metal-glass interface, along with b) transmission-mode and c) reflection-mode SNOM images as well as d) Nanoscale temperature oscillation profile [$\Delta T_{f,}(x,y)$) measured by ω-ω pump-probe near-field scanning thermoreflectance imaging. e) Thermal expansion maps at 100 Hz; f) 200 Hz; g) 400 Hz; and h) 700 Hz obtained from ω-2ω pump-probe measurements. i) Schematic representation of the sparse ML-G platelets, for which j) a negative linear thermal expansion coefficient (TEC) is observed up to 90-layer thickness. A thermomechanical transition occurs at ~175 layers, where the TEC

becomes positive, thus demonstrating the appearance of graphite-like characteristics. The negative TEC can be associated to the in-plane bending mode of graphene, whereas the positive TEC (layer-number independent at > 200 layers) can be assigned to the vibrational modes along the crystallographic z-axis of graphite.

**Table I.** Thermal Expansion Coefficient (TEC) comparison of NeFTEI for Au with existing probing methods (where XAS indicates X-Ray Absorption Spectroscopy and EXAFS indicates Extended X-Ray Absorption Fine Structure).

| Method | TEC (x $10^{-6}$ $K^{-1}$) | Conditions | Ref. |
|---|---|---|---|
| Theoretical/Simulations | 43.04 | NA | 40 |
|  | 130 |  | 41 |
|  | 35-63 |  | 42 |
|  | 14-16 |  | 43 |
| Dilatometry | 3.08 | T = 30 K | 44 |
| Stress-Strain Curves | 42.6 | T = 370 K | 45 |
|  | 0-47 | T = 0 - 550 K | 46, 47 |
| Thermal Resistance | 14.2 | NA | 48 |
| X-Ray Diffraction | 15-43 | T = 255 K | 49 |
|  | 14-18 | T = NA | 50 |
| XAS/EXAFS | 10-17 | T = 300 – 2000 K | 51 |
|  | 15 | T = 300 K | 52 |
|  | 14 | NA | 53 |
|  | 7.5 | NA | 54 |
| SNOD | $\alpha_{V,ave}$ = 51.35 ± 6.89 | T = 298 K | This Work |
|  | $\alpha_{L,ave}$ = 17.12 ± 2.30 |  |  |

**Table II.** Comparison of TEC from SNOD experiments on multi-layer graphene/graphite nanoplatelets with those achieved from macroscopic probing methods available from the literature (where EELS indicates Electron Energy Loss Spectroscopy).

| Methods | Sample | TEC (x $10^{-6}$ $K^{-1}$) | Ref. |
|---|---|---|---|
| See Ref. | Graphite | 0 - 40 | 57-59 |
| Nano EELS | Graphene/ Cu Mesh | Mono: -2.14 ± 0.79<br>Bilayer: -1.09 ± 0.25<br>Trilayer: -0.87 ± 0.17<br>Bulk: -0.07 ± 0.01 | 60 |
| Low Temperature Resonance | Au/Graphene/SiO$_2$/Si | -7.4 | 61 |
| Raman Spectroscopy | Graphene/LiNbO$_3$ | -10 to -5 | 62 |
| | Au/Graphene/ SiO$_2$/Si | Mono: -3.2 ± 0.2<br>Bilayer: -3.6 ± 0.4<br>Trilayer: -3.8 ± 0.6 | 63 |
| | Graphene/ODTS | -0.6 ± 0.5 | 64 |
| | Graphene/SiO$_2$/Si | -3.68 ± 0.49 | 65 |
| | Graphene/Ge | -(9.4±7.7) | 66 |
| Stress-Strain Curve | 3D/Graphene | -3.69 ± 0.12 to -2.71±0.28 | 67 |
| | Pris-Graphene/SiO$_2$/Si | -3.27 | 68 |
| | Graphene/SiO$_2$/Si | -7 to 2.5 | 69 |
| | Graphene/Ge | 30 | 70 |
| SNOD | Graphite/Glass | Site Specific<br>33.75 ± 3.24 | This Work |
| SNOD | ML-G/Glass | Site Specific<br>-5.77 ± 3.79 | This Work |

# Supporting Information

**A contactless scanning near-field optical dilatometer imaging the thermal expansivity of inhomogeneous 2D materials and thin films at the nanoscale.**

*Victor Wong, Sabastine Ezugwu & Giovanni Fanchini\**

V. Wong, S. Ezugwu & G. Fanchini

Department of Physics & Astronomy, University of Western Ontario, London ON, Canada N6A3K7

E-mail: gfanchin@uwo.ca

G. Fanchini

Department of Chemistry, University of Western Ontario, London ON, Canada N6A3K7

**Table of contents**

## S1. Macroscopic Thermal Expansion measurements via Optical Dilatometry

*S1.1. Optical Dilatometer for Macroscopic Measurements* - The optical dilatometer for macroscopic measurements (Figure S1a) includes i) a ceramic hot plate for sample heating (Cole-Parmer, 4-inch x 4 inch) and ii) an optical reflectometer-profilometer (Filmetrics F20) for measuring thin-film thicknesses under an illuminator through white-light interferometry (WLI) (Figure S1b).[s1,s2] Samples are placed at the center of the hot plate at room temperature where a the film thickness is measured with the reflectometer-profilometer. After the initial room-temperature scan, the sample is heated to fixed higher temperatures, at which an interferometry scan yielding the thin-film thickness is recorded again with the F20 reflectometer. Temperature at each step was measured using K-thermocouples both at the top of the film and within the hot plate with a difference of no less than 20°C, and the temperature at each step is kept stable through the feedback loop connecting the hot-plate heater with the hot-plate thermocouple. Interferometry scans offering thin-film thickness measurements are repeated for temperatures between 50°C and 540 °C, and the expansion in the thin-film thickness due to temperature yields the thermal expansion coefficient (TEC) of the thin film material. Thickness data are returned through the simulation of the interferometry scans using the Filmetrics built-in software of the F20 instrument.[s2]

*S1.2. Calibration of the Optical Dilatometer* - Calibration of our optical dilatometer was performed using commercial, 300-nm thick, $SiO_2$ layers thermally grown on a p-type silicon wafer (MTI Corp. Research Grade, B-doped, 1.455 refractive index) and is presented in **Figure S1.** The linear thermal expansion coefficient of 300 nm $SiO_2$ determined from constructive interferometry peaks A, B and C in Figure S1c are presented in Figure S1d. The origin of the shifts of peaks A,

B and C as a function of temperature can be explained by the principles of wave optics. Transmitted light will experience a phase shift ($\delta_j$, j = 0, 1, 2…., n) in its optical path due to multiple reflections (Figure S1b) in the thin-film medium prior to emerging from the planar surface. Reflected light waves from different optical paths emerging in phase with each other will undergo constructive interference and result in a reflectance peak in the measured spectrum. On the contrary, emerging reflected light from the medium that are out of phase will interfere destructively, thus resulting in a dip in the reflectometry spectra. The number of peaks in Figure S1c correspond to the number of consecutive constructive interference phase shifts due to multiple reflections measurable by our dilatometer. As the thin film is being heated, the sample is expanding which results in a change of optical path phase shifts ($\delta_j$', j = 0, 1, 2…., n). The difference in phase shift between the final and initial heating corresponds to a change in sample thickness and can be used to determine the thermal expansion coefficient of the material.

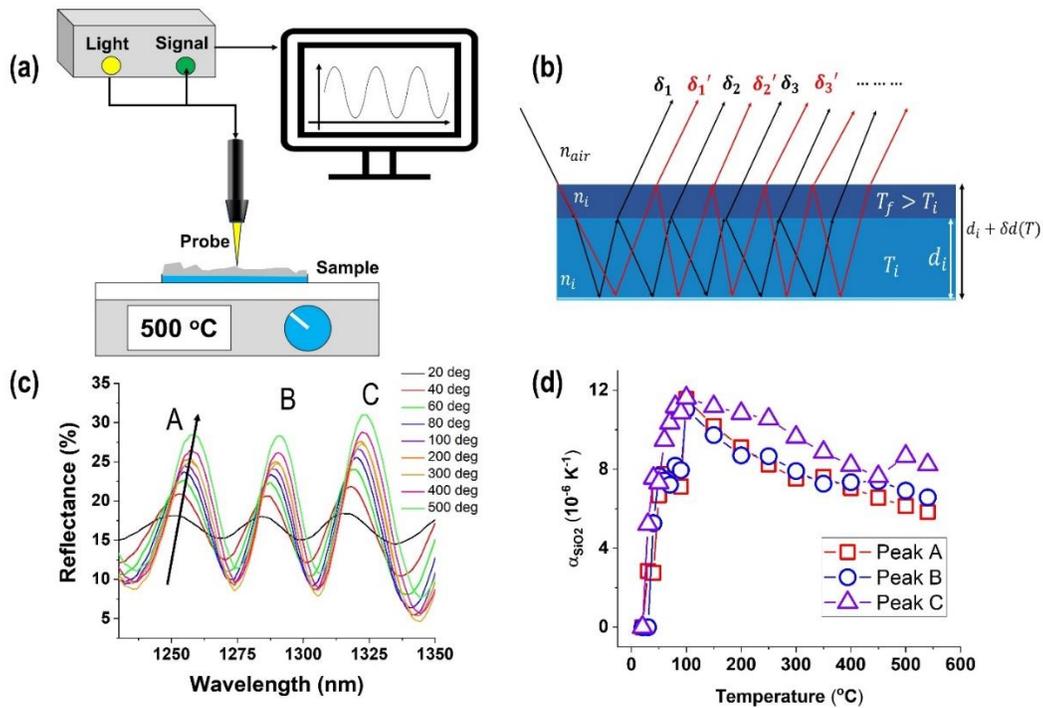

**Figure S1**. *a) Instrumental setup of the optical dilatometer for probing macroscopic thermal*

*expansion; b) schematic of the physical principle of WLI; c) thermoreflectance spectra of 300 nm SiO$_2$ film on Si as a function of temperature and d) the corresponding linear thermal expansion coefficient as a function of temperature. The spectral shift in peaks A, B, and C is a result of thermally induced expansion in the thin-film and can be used to determine the TEC.*

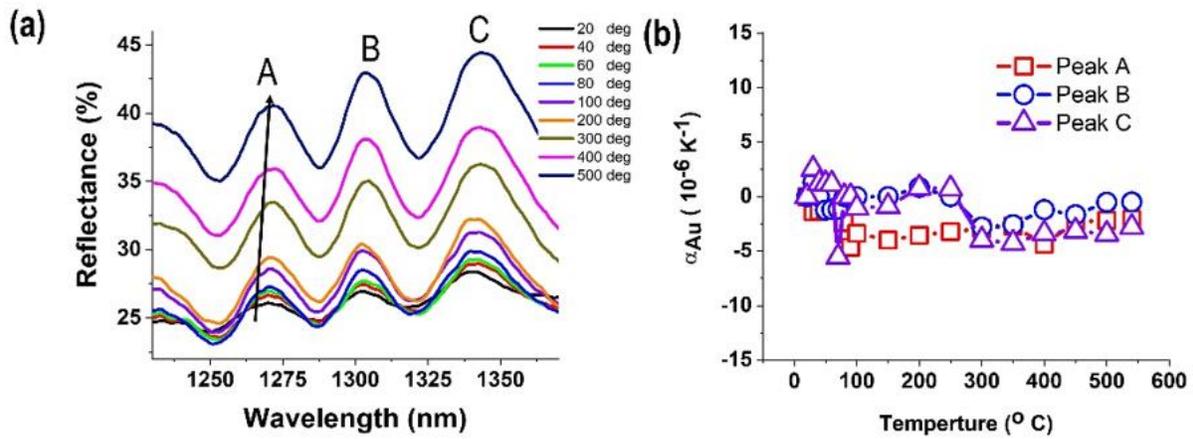

**Figure S2**. *a) Thermoreflectance spectra at varying temperatures of thermally evaporated nanogranular Au thin film on glass; and b) the corresponding linear thermal expansion coefficient as a function of temperature. The spectral shift in the reflectance spectrum is a result of thermally induced expansion in the thin-film and can be used to determine the TEC.*

*S1.3. Macroscopic Optical Dilatometry measurements in Nanogranular Au thin films* - **Figure S2** shows the temperature-varying reflectance spectrums of a nanogranular Au thin film on glass with 80 nm thickness from 20°C (no heating) to 540 °C. We can observe that there is a spectral shift in the reflectance measurements due to an increase in temperature with respect to the spectra recorded at 20°C. Panel d shows that the linear thermal expansion of Au is relatively constant with temperature dependent fluctuations ranging between ±5 x $10^{-6}$ $K^{-1}$. Irrespective of the thermal expansion trend, it should be noted that the determined thermal expansion coefficient in panel d does not yield a sensible value of Au in comparison with the results from various research groups. The discrepancy between the thermal expansion results for our Au thin film is due to the limitations of the optical dilatometer to accurately measure the dimensionality change in the film. The working principle of the optical dilatometer is based on the theory of multiple reflections which measures the consecutive interference patterns of the reflected light waves to determine the sample thickness. In the theory of multiple reflections, the sample thickness is assumed to be constant throughout the entire film, a condition that is rarely satisfied by mesoscopic structured thin films. Thus, in our macroscopic optical dilatometer the thermal expansion measurements only provides a reasonable value for a film of uniform thickness (e.g., Figure S1 for $SiO_2$) and yields highly questionable results for inhomogeneous mesoscopic thin films (e.g., our continuous Au film in this study). Therefore, there is an essential need for the development of nanoscopic thermal techniques that will provide reliable thermal expansion information to account for the inhomogeneity in mesoscopic thin films, a feature lacking in most if not all current macroscopic thermal expansion techniques.

**S2. Considering the substrate's thermal expansion in nanoscale SNOD measurements**

*S2.1. General Considerations -* In this section, we present the model to consider the role of the substrate in determining the surface thermal expansivity in eq (5). Two component are considered and superimposed as described by eq (6):

$$\Delta T_s(x_0, y_0, z_0) = T_{s,av}(z_0) + \delta T_s(x_0, y_0, z_0). \tag{s1}$$

The two parts to be linearly superimposed are as follows:

i) The average substrate temperature over the entire measured region [$T_{s,av}(z_0)$] which arise from solving the one-dimensional heat equation (fig. 1c, top):

$$D_s \frac{\partial^2 T_{s,av}(z_0,t)}{\partial z_0^2} = \frac{\partial T_{s,av}(z_0,t)}{\partial t} \quad , \tag{s2}$$

with boundary conditions

$$T_{s,av}(z_0 = 0, t) = T_{f,av} \cos(\omega t) \tag{s3}$$

ii) The local fluctuations of surface temperature at a point $(x,y) \in A$, where A is centered at the measured point $(x_0, y_0)$ which lead to the 3D heat equation (fig. 3, bottom):

$$D_s \left[ \frac{\partial^2 \delta\Theta_s(r,\varphi,t)}{\partial r^2} + \frac{2}{r}\frac{\partial \delta\Theta_s(r,\varphi,t)}{\partial r} + \frac{1}{r^2}\frac{\partial^2 \delta\Theta_s(r,\varphi,t)}{\partial \varphi^2} + \frac{\cot(\varphi)}{r^2}\frac{\partial T_s(r,\varphi,t)}{\partial \varphi} \right] = \frac{\partial \delta\Theta_s(r,\varphi,t)}{\partial t} \quad , \tag{s4}$$

with boundary conditions

$$\delta\Theta_s(x_0, y_0, z_0=0, t) = \delta T_{f,av} \cos(\omega t) \, \delta A \tag{s5}$$

where $\delta A$ is the area of the pixel scanned by the AFM/SNOM system (i.e., if a 50x50μm area was scanned at 256x/256 pixel, the area of the individual pixel would be $\delta A = (50/256)$x$(50/256) = 195$ x $195$ nm². Eq. (s4) has already been written by considering the symmetry over the angular component $\theta$ in the homogeneous substrate, and posing $r = [(x-x_0)^2 + (y-y_0)^2 + z_0^2]^{1/2}$ and $\varphi = \operatorname{atan}\{z_0/[(x-x_0)^2+(y-y_0)^2]^{1/2}\}$.

Also, in eqs (s2) and (s4), $D_s$ is the thermal diffusivity of the substrate, which is assumed, for the sake of simplicity, to be homogeneous (even though this assumption could be relaxed through a more sophisticated numerical model). Eq (s2) is readily solved as

$$T_{s,av}(z_0, t) = T_{f,av} \exp[-(\omega/D_s)^{1/2} z_0] \cos(\omega t), \tag{s6}$$

irrespectively of the scanned point $(x_0, y_0)$. This happens because the uniform heat distribution assumed by eq (s2) indicates that each point receives, from points at the same level $z_0$, as much heat as it dissipates, so the net heat flux is zero. We will now thus focus specifically on eq (s4).

*S2.2. Considering the Effect of Local Thermal Fluctuations on the Substrate -* Eq (s4) can be solved, at any heated point ($x_0$, $y_0$) by separation of variables, as

$$\delta\Theta_s(x_0, y_0, z_0=0, t) = R(r)\, \Phi(\varphi)\, f(t), \tag{s7}$$

where R(r), Φ(φ), and f(t) express the radial, azimuthal and time components, respectively, as demonstrated in Figure S3.

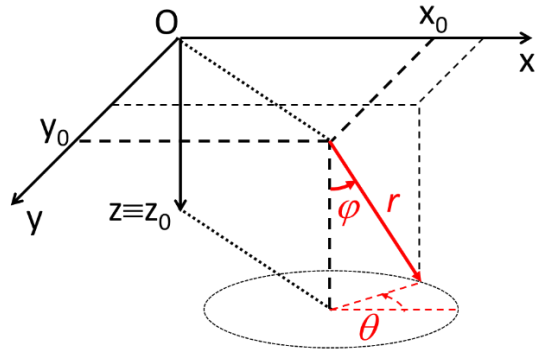

**Figure S3.** *Configuration and variables used for the analytical solution of heat equation [eq (s4)] to consider impact on the substrate from the local thermal fluctuations from the inhomogeneous heating of the superimposed thin film and/or 2D material.*

Hence, eq (s4) can be separated into three ordinary differential eigenvalue equations, respectively in r, φ, and t, which will take the forms:

$$df(t)/dt = \Omega f(t) \tag{s8}$$

$$\frac{d^2\Phi(\varphi)}{d\varphi^2} + \cot(\varphi)\frac{d\Phi(\varphi)}{d\varphi} = -L\Phi(\varphi) \tag{s9}$$

$$r^2\frac{d^2R(r)}{dr^2} + 2r\frac{dR(r)}{dr} - \frac{i\Omega}{D_s}r^2 R(r) = LR(r), \tag{s10}$$

where $\Omega$ and $\pm L$ are the corresponding eigenvalues. As far as our boundary conditions are concerned, eq (s8) requires oscillatory solutions at $\Omega = 0$ and $\pm i\omega$, while eq (s9) is a Legendre-type equation with solutions for eigenvalues

$$L = l(l+1) \qquad (l = 0, +1, ..., +\infty) \qquad (s11)$$

which also constrain the eigenvalues of eq (s10). The most general solution of (s4) has the form

$$\delta\Theta_s(r\varphi t) = [1 + \cos(\omega t)] P_l(\cos\varphi) \sum_{l=0}^{\infty} [b_l j_l(\sqrt{i\omega/D_s}\, r) + c_l y_l(\sqrt{i\omega/D_s}\, r)] \qquad (s12)$$

where $P_l(\cos\varphi)$ are associated $l$-th order Legendre polynomials and $a_l$ and $b_l$ are suitable projections of each spherical Bessel function, $j_l$ and $y_l$, which are determined by the boundary conditions (s5):

(s13)

$$b_l = \frac{2\pi \delta T_f(x_0, y_0)}{\delta A} \int_{\frac{\pi}{2}}^{\frac{\pi}{2}} P_l^*(\cos\varphi) \sin\varphi\, d\varphi \int_{-\infty}^{+\infty} j_l^*(\sqrt{i\omega/D_s}\, r) r^2 dr$$

$$c_l = \frac{2\pi \delta T_f(x_0, y_0)}{\delta A} \int_0^{\frac{\pi}{2}} P_l^*(\cos\varphi) \sin\varphi\, d\varphi \int_{-\infty}^{+\infty} y_l^*(\sqrt{i\omega/D_s}\, r) r^2 dr \qquad (s14)$$

Calculation of these projections can be performed through integration in the complex space using residue's theorem, [s3] which leads to

$$c_l = 0 \qquad (s15)$$

$$b_l = -2\delta T_f(x_0, y_0)/\delta A \quad \text{and} \quad b_l = 0 \quad (l \neq -1). \qquad (s16)$$

This result could be expected because the Legendre polynomial $P_1(\cos\varphi) = \cos\varphi$ is the only one preserving the symmetry of the heated system leading to an isotropic distribution of heat within the substrate. Therefore, only the first-order harmonic in $y_l$ contributes to the temperature profile within the substrate, and to thermal expansion.

By replacing eqs (s15) and (s16) into (s12) and by considering that [s4]

$$j_l[(i\omega/D_s)^{1/2} r] = \sin[(\omega/D_s)^{1/2} r]/[(\omega/D_s)^{1/2} r] \qquad (s17)$$

we obtain the following expression for the substrate temperature from heating at point ($x_0$ $y_0$):

$$\delta\Theta_s(x-x_0, y-y_0, z_0, t) = \frac{2\delta T_f(x_0,y_0)\sin(r\sqrt{i\omega/D_s}\,r)}{\delta A\sqrt{\omega/D_s}\,r}\cos\varphi\,[1+\cos(\omega t)] \qquad (s18)$$

where $r = [(x-x_0)^2 + (y-y_0)^2 + z_0^2]^{1/2}$ and $\varphi = \text{atan}\{z_0/[(x-x_0)^2+(y-y_0)^2]^{1/2}\}$. Although $\delta\Theta_s(x_0, y_0, z_0=0, t)$ appears to diverge at $r = 0$, this is just apparent, because also $\delta A$ tends to zero in a continuous approximation. By replacing eq (s18) into eq 7, the effect on the substrate of local temperature fluctuations at the film on the surface can be properly taken into account, and a suitable cut-off can be applied in eq 7 to determine the region of points (x,y) contributing to the integral in that equation.